

\hsize=6.5truein
\hoffset=.3truein
\vsize=8.9truein
\voffset=.1truein
\font\twelverm=cmr10 scaled 1200    \font\twelvei=cmmi10 scaled 1200
\font\twelvesy=cmsy10 scaled 1200   \font\twelveex=cmex10 scaled 1200
\font\twelvebf=cmbx10 scaled 1200   \font\twelvesl=cmsl10 scaled 1200
\font\twelvett=cmtt10 scaled 1200   \font\twelveit=cmti10 scaled 1200
\skewchar\twelvei='177   \skewchar\twelvesy='60
\def\twelvepoint{\normalbaselineskip=14pt
  \abovedisplayskip 12.4pt plus 3pt minus 9pt
  \belowdisplayskip 12.4pt plus 3pt minus 9pt
  \abovedisplayshortskip 0pt plus 3pt
  \belowdisplayshortskip 7.2pt plus 3pt minus 4pt
  \smallskipamount=3.6pt plus1.2pt minus1.2pt
  \medskipamount=7.2pt plus2.4pt minus2.4pt
  \bigskipamount=14.4pt plus4.8pt minus4.8pt
  \def\rm{\fam0\twelverm}          \def\it{\fam\itfam\twelveit}%
  \def\sl{\fam\slfam\twelvesl}     \def\bf{\fam\bffam\twelvebf}%
  \def\mit{\fam 1}                 \def\cal{\fam 2}%
  \def\tt{\twelvett}
  \textfont0=\twelverm   \scriptfont0=\tenrm   \scriptscriptfont0=\sevenrm
  \textfont1=\twelvei    \scriptfont1=\teni    \scriptscriptfont1=\seveni
  \textfont2=\twelvesy   \scriptfont2=\tensy   \scriptscriptfont2=\sevensy
  \textfont3=\twelveex   \scriptfont3=\twelveex  \scriptscriptfont3=\twelveex
  \textfont\itfam=\twelveit
  \textfont\slfam=\twelvesl
  \textfont\bffam=\twelvebf \scriptfont\bffam=\tenbf
  \scriptscriptfont\bffam=\sevenbf
  \normalbaselines\rm}

\def\beginlinemode{\endmode
  \begingroup\parskip=0pt \obeylines\def\\{\par}\def\endmode{\par\endgroup}}
\def\beginparmode{\endmode
  \begingroup \def\endmode{\par\endgroup}}
\let\endmode=\par
{\obeylines\gdef\
{}}
\def\singlespace{\baselineskip=\normalbaselineskip}
\def\oneandahalfspace{\baselineskip=\normalbaselineskip
  \multiply\baselineskip by 3 \divide\baselineskip by 2}
\def\doublespace{\baselineskip=\normalbaselineskip \multiply\baselineskip by 2}
\newcount\firstpageno
\firstpageno=2
\footline={\ifnum\pageno<\firstpageno{\hfil}\else{\hfil\twelverm\folio\hfil}\fi}
\let\rawfootnote=\footnote              
\def\footnote#1#2{{\rm\singlespace\parindent=0pt\rawfootnote{#1}{#2}}}
\def\raggedcenter{\leftskip=2em plus 12em \rightskip=\leftskip
  \parindent=0pt \parfillskip=0pt \spaceskip=.3333em \xspaceskip=.5em
  \pretolerance=9999 \tolerance=9999
  \hyphenpenalty=9999 \exhyphenpenalty=9999 }
\parskip=\medskipamount
\twelvepoint            
\overfullrule=0pt       
\def\preprintno#1{
 \rightline{\rm #1}}    
\def\author                     
  {\vskip 3pt plus 0.2fill \beginlinemode
   \singlespace \raggedcenter \twelvesc}
\def\affil                      
  {\vskip 3pt plus 0.1fill \beginlinemode
   \oneandahalfspace \raggedcenter \sl}
\def\abstract                   
  {\vskip 3pt plus 0.3fill \beginparmode
   \doublespace \narrower \noindent ABSTRACT: }
\def\endtitlepage               
  {\endpage                     
   \body}
\def\body                       
  {\beginparmode}               

\def\subhead#1{                 
  \vskip 0.1truein             
  {\raggedcenter #1 \par}
   \nobreak\vskip 0.1truein\nobreak}
\def\refto#1{$|{#1}$}           
\def\references                 
  {\subhead{References}         
   \beginparmode
   \frenchspacing \parindent=0pt \leftskip=1truecm
   \parskip=8pt plus 3pt \everypar{\hangindent=\parindent}}
\gdef\refis#1{\indent\hbox to 0pt{\hss#1.~}}    
\gdef\journal#1, #2, #3, 1#4#5#6{               
    {\sl #1~}{\bf #2}, #3, (1#4#5#6)}           
\def\refstylenp{                
  \gdef\refto##1{$^{(##1)}$}                                
[]
  \gdef\refis##1{\indent\hbox to 0pt{\hss##1)~}}        
  \gdef\journal##1, ##2, ##3, ##4 {                     
     {\sl ##1~}{\bf ##2~}(##3) ##4 }}
\def\refstyleprnp{              
  \gdef\refto##1{$^{(##1)}$}                                
[]
  \gdef\refis##1{\indent\hbox to 0pt{\hss##1)~}}        
  \gdef\journal##1, ##2, ##3, 1##4##5##6{               
    {\sl ##1~}{\bf ##2~}(1##4##5##6) ##3}}

\def\pr{\journal Phys. Rev., }

\def\prl{\journal Phys. Rev. Lett., }
\def\prpts{\journal Phys. Rep., }
\def\np{\journal Nucl. Phys., }
\def\pl{\journal Phys. Lett., }

\def\ptp{\journal Prog Theo Physics, }
\def\endreferences{\body}
\def\endpage                    
  {\vfill\eject}
\def\endpaper                   
  {\endmode\vfill\supereject}
\def\endit
  {\endpaper\end}
\def\ref#1{Ref. #1}                     
\def\Ref#1{Ref. #1}                     

\def\m@th{\mathsurround=0pt }
\font\twelvesc=cmcsc10 scaled 1200
\def\cite#1{{#1}}
\def\(#1){(\call{#1})}
\def\call#1{{#1}}
\def\taghead#1{}
\def\leaderfill{\leaders\hbox to 1em{\hss.\hss}\hfill}
\def\twiddle{\lower.9ex\rlap{$\kern-.1em\scriptstyle\sim$}}
\def\bigtwiddle{\lower1.ex\rlap{$\sim$}}
\def\gtwid{\mathrel{\raise.3ex\hbox{$>$\kern-.75em\lower1ex\hbox{$\sim$}}}}
\def\ltwid{\mathrel{\raise.3ex\hbox{$<$\kern-.75em\lower1ex\hbox{$\sim$}}}}
\def\square{\kern1pt\vbox{\hrule height 1.2pt\hbox{\vrule width 1.2pt\hskip 3pt
   \vbox{\vskip 6pt}\hskip 3pt\vrule width 0.6pt}\hrule height 0.6pt}\kern1pt}
\catcode`@=11
\newcount\tagnumber\tagnumber=0

\immediate\newwrite\eqnfile
\newif\if@qnfile\@qnfilefalse
\def\write@qn#1{}
\def\writenew@qn#1{}
\def\w@rnwrite#1{\write@qn{#1}\message{#1}}
\def\@rrwrite#1{\write@qn{#1}\errmessage{#1}}

\def\taghead#1{\gdef\t@ghead{#1}\global\tagnumber=0}
\def\t@ghead{}

\expandafter\def\csname @qnnum-3\endcsname
  {{\t@ghead\advance\tagnumber by -3\relax\number\tagnumber}}
\expandafter\def\csname @qnnum-2\endcsname
  {{\t@ghead\advance\tagnumber by -2\relax\number\tagnumber}}
\expandafter\def\csname @qnnum-1\endcsname
  {{\t@ghead\advance\tagnumber by -1\relax\number\tagnumber}}
\expandafter\def\csname @qnnum0\endcsname
  {\t@ghead\number\tagnumber}
\expandafter\def\csname @qnnum+1\endcsname
  {{\t@ghead\advance\tagnumber by 1\relax\number\tagnumber}}
\expandafter\def\csname @qnnum+2\endcsname
  {{\t@ghead\advance\tagnumber by 2\relax\number\tagnumber}}
\expandafter\def\csname @qnnum+3\endcsname
  {{\t@ghead\advance\tagnumber by 3\relax\number\tagnumber}}

\def\equationfile{%
  \@qnfiletrue\immediate\openout\eqnfile=\jobname.eqn%
  \def\write@qn##1{\if@qnfile\immediate\write\eqnfile{##1}\fi}
  \def\writenew@qn##1{\if@qnfile\immediate\write\eqnfile
    {\noexpand\tag{##1} = (\t@ghead\number\tagnumber)}\fi}
}

\def\callall#1{\xdef#1##1{#1{\noexpand\call{##1}}}}
\def\call#1{\each@rg\callr@nge{#1}}

\def\each@rg#1#2{{\let\thecsname=#1\expandafter\first@rg#2,\end,}}
\def\first@rg#1,{\thecsname{#1}\apply@rg}
\def\apply@rg#1,{\ifx\end#1\let\next=\relax%
\else,\thecsname{#1}\let\next=\apply@rg\fi\next}

\def\callr@nge#1{\calldor@nge#1-\end-}
\def\callr@ngeat#1\end-{#1}
\def\calldor@nge#1-#2-{\ifx\end#2\@qneatspace#1 %
  \else\calll@@p{#1}{#2}\callr@ngeat\fi}
\def\calll@@p#1#2{\ifnum#1>#2{\@rrwrite{Equation range #1-#2\space is bad.}
\errhelp{If you call a series of equations by the notation M-N, then M and
N must be integers, and N must be greater than or equal to M.}}\else%
 {\count0=#1\count1=#2\advance\count1
by1\relax\expandafter\@qncall\the\count0,%
  \loop\advance\count0 by1\relax%
    \ifnum\count0<\count1,\expandafter\@qncall\the\count0,%
  \repeat}\fi}

\def\@qneatspace#1#2 {\@qncall#1#2,}
\def\@qncall#1,{\ifunc@lled{#1}{\def\next{#1}\ifx\next\empty\else
  \w@rnwrite{Equation number \noexpand\(>>#1<<) has not been defined yet.}
  >>#1<<\fi}\else\csname @qnnum#1\endcsname\fi}

\let\eqnono=\eqno
\def\eqno(#1){\tag#1}
\def\tag#1$${\eqnono(\displayt@g#1 )$$}

\def\aligntag#1\endaligntag
  $${\gdef\tag##1\\{&(##1 )\cr}\eqalignno{#1\\}$$
  \gdef\tag##1$${\eqnono(\displayt@g##1 )$$}}

\def\eqalignno#1{\displ@y \tabskip\centering
  \halign to\displaywidth{\hfil$\displaystyle{##}$\tabskip\z@skip
    &$\displaystyle{{}##}$\hfil\tabskip\centering
    &\llap{$\displayt@gpar##$}\tabskip\z@skip\crcr
    #1\crcr}}

\def\displayt@gpar(#1){(\displayt@g#1 )}

\def\displayt@g#1 {\rm\ifunc@lled{#1}\global\advance\tagnumber by1
        {\def\next{#1}\ifx\next\empty\else\expandafter
        \xdef\csname @qnnum#1\endcsname{\t@ghead\number\tagnumber}\fi}%
  \writenew@qn{#1}\t@ghead\number\tagnumber\else
        {\edef\next{\t@ghead\number\tagnumber}%
        \expandafter\ifx\csname @qnnum#1\endcsname\next\else
        \w@rnwrite{Equation \noexpand\tag{#1} is a duplicate number.}\fi}%
  \csname @qnnum#1\endcsname\fi}

\def\ifunc@lled#1{\expandafter\ifx\csname @qnnum#1\endcsname\relax}

\let\@qnend=\end\gdef\end{\if@qnfile
\immediate\write16{Equation numbers written on []\jobname.EQN.}\fi\@qnend}

\catcode`@=12
\refstyleprnp
\catcode`@=11
\newcount\r@fcount \r@fcount=0
\def\refreset{\global\r@fcount=0}
\newcount\r@fcurr
\immediate\newwrite\reffile
\newif\ifr@ffile\r@ffilefalse
\def\w@rnwrite#1{\ifr@ffile\immediate\write\reffile{#1}\fi\message{#1}}

\def\writer@f#1>>{}
\def\referencefile{
  \r@ffiletrue\immediate\openout\reffile=\jobname.ref%
  \def\writer@f##1>>{\ifr@ffile\immediate\write\reffile%
    {\noexpand\refis{##1} = \csname r@fnum##1\endcsname = %
     \expandafter\expandafter\expandafter\strip@t\expandafter%
     \meaning\csname r@ftext\csname r@fnum##1\endcsname\endcsname}\fi}%
  \def\strip@t##1>>{}}

\def\citeall#1{\xdef#1##1{#1{\noexpand\cite{##1}}}}
\def\cite#1{\each@rg\citer@nge{#1}}	

\def\each@rg#1#2{{\let\thecsname=#1\expandafter\first@rg#2,\end,}}
\def\first@rg#1,{\thecsname{#1}\apply@rg}	
\def\apply@rg#1,{\ifx\end#1\let\next=\relax
\else,\thecsname{#1}\let\next=\apply@rg\fi\next}

\def\citer@nge#1{\citedor@nge#1-\end-}	
\def\citer@ngeat#1\end-{#1}
\def\citedor@nge#1-#2-{\ifx\end#2\r@featspace#1 
  \else\citel@@p{#1}{#2}\citer@ngeat\fi}	
\def\citel@@p#1#2{\ifnum#1>#2{\errmessage{Reference range #1-#2\space is bad.}%
    \errhelp{If you cite a series of references by the notation M-N, then M and
    N must be integers, and N must be greater than or equal to M.}}\else%
 {\count0=#1\count1=#2\advance\count1
by1\relax\expandafter\r@fcite\the\count0,%
  \loop\advance\count0 by1\relax
    \ifnum\count0<\count1,\expandafter\r@fcite\the\count0,%
  \repeat}\fi}

\def\r@featspace#1#2 {\r@fcite#1#2,}	
\def\r@fcite#1,{\ifuncit@d{#1}
    \newr@f{#1}%
    \expandafter\gdef\csname r@ftext\number\r@fcount\endcsname%
                     {\message{Reference #1 to be supplied.}%
                      \writer@f#1>>#1 to be supplied.\par}%
 \fi%
 \csname r@fnum#1\endcsname}
\def\ifuncit@d#1{\expandafter\ifx\csname r@fnum#1\endcsname\relax}%
\def\newr@f#1{\global\advance\r@fcount by1%
    \expandafter\xdef\csname r@fnum#1\endcsname{\number\r@fcount}}

\let\r@fis=\refis			
\def\refis#1#2#3\par{\ifuncit@d{#1}
   \newr@f{#1}%
   \w@rnwrite{Reference #1=\number\r@fcount\space is not cited up to now.}\fi%
  \expandafter\gdef\csname r@ftext\csname r@fnum#1\endcsname\endcsname%
  {\writer@f#1>>#2#3\par}}

\def\ignoreuncited{
   \def\refis##1##2##3\par{\ifuncit@d{##1}%
     \else\expandafter\gdef\csname r@ftext\csname
r@fnum##1\endcsname\endcsname%
     {\writer@f##1>>##2##3\par}\fi}}

\def\r@ferr{\endreferences\errmessage{I was expecting to see
\noexpand\endreferences before now;  I have inserted it here.}}
\let\r@ferences=\references
\def\references{\r@ferences\def\endmode{\r@ferr\par\endgroup}}

\let\endr@ferences=\endreferences
\def\endreferences{\r@fcurr=0
  {\loop\ifnum\r@fcurr<\r@fcount
    \advance\r@fcurr by 1\relax\expandafter\r@fis\expandafter{\number\r@fcurr}%
    \csname r@ftext\number\r@fcurr\endcsname%
  \repeat}\gdef\r@ferr{}\global\r@fcount=0\endr@ferences}

\let\r@fend=\endpaper\gdef\endpaper{\ifr@ffile
\immediate\write16{Cross References written on []\jobname.REF.}\fi\r@fend}

\catcode`@=12

\citeall\refto		
\citeall\ref		%
\citeall\Ref		%

\referencefile

\def\frac#1/#2{#1 / #2}
\def\uof{Department of Physics\\University of Florida\\Gainesville FL 32611}

\def\oneandthreefifthsspace{\baselineskip=\normalbaselineskip
  \multiply\baselineskip by 8 \divide\baselineskip by 5}

\font\titlefont=cmr10 scaled\magstep3
\def\bigtitle                      
  {\null\vskip 3pt plus 0.2fill
   \beginlinemode \doublespace \raggedcenter \titlefont}

\def\uof{Institute for Fundamental Theory\\Department of Physics,
University of Florida\\Gainesville FL 32611}
\preprintno{UFIFT-HEP-95-7}
\bigtitle{Consequences of an Abelian Family Symmetry}
 \author Pierre Ramond\affil\uof
\vskip .5cm
\centerline{Contribution to the 25th Anniversary Volume}
\centerline{Centre de Recherches Math\'ematiques, Universit\'e de
Montr\'eal}
\body
\abstract
The addition of an Abelian family symmetry to
the Minimal Supersymmetric Standard Model reproduces the observed
hierarchies of quark and lepton masses and
quark mixing angles, only if it is anomalous.  Green-Schwarz
compensation of its anomalies requires the electroweak mixing angle to
be $\sin^2\theta_w=3/8$ at the string scale, without any assumed GUT
structure, suggesting a superstring origin for the standard model.
The analysis is extended to neutrino masses and the lepton mixing
matrix.
\endtitlepage
\oneandthreefifthsspace

\subhead{1.~\bf Introduction}
\taghead{1.}

The relative complexity of the standard model of strong and electroweak
interactions suggests that it is the low energy manifestation of more
fundamental theory. There are few hints as to the nature of this theory
or value of the scale at which it becomes operative.

These questions can be studied in the context of the   extension of the
standard model to $N=1$ supersymmetry[\cite{reviews}] which allows for
its perturbative extrapolation to near Planckian scales, where the gauge
couplings [\cite{unification}]and some Yukawa couplings[\cite{btau}]
appear to converge. This raises the hope that the $N=1$ standard model
at short distances is much simpler than at experimental scales, although
we do not seem to have sufficient information to determine exactly the
type of structure it describes, a GUT theory[\cite{gut}], or a direct
descendant of superstrings. Fortunately there is more information, as
not all parameters are the same at the unification scale, suggesting in
fact a strong hierarchy among the masses of the quarks and the leptons,
indicated by the orders of magnitude estimates[\cite{RRR}]

$${m_u\over
m_t}={\cal O}(\lambda^8)\ ;\qquad {m_c\over m_t}={\cal O}(\lambda^4)\ ;
\eqno(top)$$ $${m_d\over m_b}={\cal O}(\lambda^4)\ ; \qquad {m_s\over
m_b}={\cal O}(\lambda^2)\ ,\eqno(bottom)$$
where, following
Wolfenstein's parametrization[\cite{wolf}], we use the Cabibbo angle
$\lambda$, as expansion parameter. The charged lepton masses also
satisfy similar relations
$${m_e\over m_\tau}={\cal O}(\lambda^4)\ ;
\qquad {m_\mu\over m_\tau}={\cal O}(\lambda^2)\ .\eqno(lept)$$
The mass
hierarchy appears to be geometrical in each sector. The equality
$$m_b=m_\tau\ ,$$
known to be valid in the ultraviolet[\cite{btau}],
yields the estimate
$${m_dm_sm_b\over m_e m_\mu m_\tau}={\cal O}(1) \
.\eqno(goodrat)$$
We assume that the mechanism which sets these orders
of magnitude is  an Abelian family symmetry, in the manner originally
suggested by Froggatt and Nielsen[\cite{FN}].

In the following, we  suggest that these orders of magnitude are
determined by a gauged family Abelian symmetry. This is an old idea,
some aspects of which has been revisited in the recent literature
[\cite{LNS, IR,PAPAG}]. The work presented below is the result of a
collaboration with P. Bin\' etruy and S. Lavignac[\cite{BR,BLR}].
There are also closely related work in the recent literature,
but with different emphases[\cite{JS,DPS,NIR}].

Our framework is the minimal extension of the Standard Model to $N=1$
supersymmetry, including the so-called $\mu$ term, $P=\mu H_uH_d\ .$
The most remarkable conclusion is that to reproduce these estimates,
the symmetry must be anomalous. The anomalies can be compensated by
the Green-Schwarz mechanism, fixing[\cite{Ib}]
$\sin^2\theta_w=3/8$, in perfect agreement with data when
extrapolated to the infrared.

\subhead{2.~\bf Quark Masses and Mixing Angles}
\taghead{2.}
The most general Abelian charge assignments  to
the particles of the Supersymmetric Standard Model can be written as
$$X=X^{}_0+X^{}_3+{\sqrt 3}X^{}_8\ ,\eqno(X)$$
where $X_0$ is the family-independent part, $X_3$ is along $\lambda_3$,
and $X_8$ is along $\lambda_8$, the two diagonal Gell-Mann matrices of
the $SU(3)$ family space in each charge sector. In a basis where the
entries correspond to the components in the family space of the fields
${\bf Q}$, $\overline{\bf u}$, $\overline{\bf d}$, $L$, and $\overline
e$, we can write the different components in the form
$$X_i^{}=(a^{}_i,b^{}_i,c_i^{},d_i^{},e_i^{})\ ,\eqno(Xfamily)$$
for $i=0,3,8$.
The Higgs doublets $H_{u,d}$ have zero X-charge because of the $\mu$
term.

The tree-level Yukawa coupling involves {\it only}
the third family (implicitly choosing the third direction in family
space),
$$y_t{\bf Q}_3\overline{\bf u}_3H_u+y_b{\bf Q}_3\overline{\bf d}_3H_d+
y_\tau L_3\overline e_3H_d\ ,\eqno(Yuk)$$
where the $y_i$'s are the Yukawa couplings. This generates the relations
$$a_0^{}+b_0^{}=2(a_8+b_8)\ ,\qquad a_0^{}+c_0^{}=2(a_8+c_8)\ ,\qquad
d_0^{}+e_0^{}=2(d_8+e_8)\ .$$
The other elements of the Yukawa matrices are zero at tree-level
because of X-charge  conservation.  Let $x_{ij}$
be the  excess X-charges at each of their entries; for the charge 2/3
Yukawa matrix they are

$$
\pmatrix{3(a_8+b_8)+a_3+b_3&3(a_8+b_8)+a_3-b_3&3a_8+a_3\cr
3(a_8+b_8)-a_3+b_3&
3(a_8+b_8)-a_3-b_3&3a_8-a_3\cr
3b_8+b_3&3b_8-b_3&0\cr}\ .\eqno(Yuky)$$

In the charge $-1/3$ sector, the $b_i$ are replaced by the $c_i$, and in
the charge $-1$ sector, $a_i,b_i$ are replaced by $d_i,e_i$,
respectively.

Introduce an electroweak singlet field $\theta$  with X-charge
$-x$, to soak up the excess charge at each entry, $x_{ij}$, yielding
an interaction of higher dimensions with no hypercharge [\cite{FN,hall}]

$${\bf Q}_i\overline{\bf u}_jH_u\left({\theta\over M_u}\right)^{n_{ij}}\ ,
\eqno(nonren)$$
where  the $n_{ij}$ are positive numbers which satisfy
$$x_{ij}-xn_{ij}= 0 \ ,\eqno(conserve)$$
and $M_u$ is some large scale.   In a perturbative framework, the $n_{ij}$
are expected to be integers.
With only one field $\theta$, not chaperoned by its vectorlike
partner, invariance under supersymmetry then naturally[\cite{LNS}]
generates a true texture zero whenever a Yukawa matrix element has
negative excess $X$-charge in units of (-$x$), and non-zero entries
correspond only to positive excess X-charge. Henceforth we normalize X
so that $x=1$.

We assume that the electroweak singlet  $\theta$ develops a vacuum
expectation value smaller than $M_u$,
 producing a {\it small parameter}, $\lambda_u\sim\theta/M_u$. This is
what happens in many compactified superstring theories[\cite{break}].
The $n_{ij}$
then determine the order of magnitude of the entries in the Yukawa
matrices[\cite{FN}]. It may seem that because the  masses $M_u$,
and thus the expansion
parameters are  in principle different in the three charge sectors, we
are introducing many unknowns in our description, but most of the
 conclusions we can reach with this simple assumption
depend only on the {\it existence} of these small parameters, {\it not on
their values}. Also, since the down quark and lepton sectors share
the same electroweak quantum numbers, we expect them to be the same
for the charge -1 and -1/3 matrices.

In the following we restrict ourselves to the case where all the
excess charges in each Yukawa matrix are positive. We leave to future
work the study of the cases where some of the excess charges are
negative, which creates a true zero in that matrix element[\cite{LNS}].
The charge $2/3$  Yukawa matrix is
$${\bf Y}_{uij}={\cal O}(\lambda_u^{n_{ij}})\ ,\eqno(oom)$$
normalized to the top quark mass. It is not hard
to diagonalize this matrix, setting
$${\bf Y}^{}_u=U^{}_uD^{}_uV_u^\dagger\ ,\eqno(diag)$$
where
$$D^{}_u={\rm diag}~( {\cal O}(\lambda_u^{3(a_8+b_8)+a_3+b_3}),{\cal
O}(\lambda_u^{3(a_8+b_8)-a_3-b_3}),{\cal O}(1))\ ,$$
and the unitary matrix $U_u$ is given by
$$U^{}_u=\pmatrix{{\cal O}(1)&{\cal O}(\lambda_u^{2a_3})&
{\cal O}(\lambda_u^{3a_8+a_3})\cr
{\cal O}(\lambda_u^{2a_3})&{\cal O}(1)&{\cal
O}(\lambda_u^{3a_8-a_3})\cr
{\cal O}(\lambda_u^{3a_8+a_3})&{\cal O}(\lambda_u^{3a_8-a_3})
&{\cal O}(1)}\ ,\eqno(udiag)$$
These are valid for a range of parameters such that
$$3a_8+3b_8>a_3+b_3>0\ .$$
We have a similar relation in the down quark sector, with the $b_i$
replaced by $c_i$. It follows that the
orders of magnitude of $U_u$ and $U_d$ are the same, but the expansion
coefficients might be different. Let us set
$\lambda^{}_u=\lambda_d^y$, with $y>0$. If $y>1$,
the orders of magnitude of
the entries of the CKM matrix are
$${\cal U}_{CKM}=\pmatrix{{\cal O}(1)&{\cal
O}(\lambda_d^{2a_3})&
{\cal O}(\lambda_d^{3a_8+a_3})\cr
{\cal O}(\lambda_d^{2a_3})&{\cal O}(1)&{\cal O}(\lambda_d^{
3a_8-a_3})\cr
{\cal O}(\lambda_d^{3a_8+a_3})&{\cal O}(\lambda_d^{
3a_8-a_3})&{\cal O}(1)}\
.\eqno(CKM)$$
If $y<1$, the expansion parameter in the above is replaced by
$\lambda_u$, that is its exponents all are multiplied by $y$. In
either case the exponents satisfy the sum rule
$$n_{12}=n_{13}-n_{23}\ ,\eqno(sr)$$
which implies that
$${V_{us}~V_{cb}\over V_{ub}}={\cal O}(1)\ ,\eqno(bbb)$$
in agreement with data (the right hand side is $\approx 3$,
and the Wolfenstein parametrization ($n_{12}=1,~n_{13}=3,~n_{23}=2$).
We note the relation between our expansion parameters with the Cabibbo angle
$$\lambda\equiv V_{us}=\lambda_{u,d}^{~~2a_3}\ ,\eqno(cabib)$$
depending on the relative magnitudes of $\lambda_u$ and $\lambda_d$.
The eigenvalue order of magnitude estimates are

$${m_u\over m_t}={\cal O}(\lambda_u^{3(a_8+b_8)+a_3+b_3})\ ;
\qquad {m_c\over m_t}={\cal
O}(\lambda_u^{3(a_8+b_8)-a_3-b_3})\ ; $$
$${m_d\over m_b}={\cal
O}(\lambda_d^{3(a_8+c_8)+a_3+c_3})\ ; \qquad {m_s\over m_b}={\cal
O}(\lambda_d^{3(a_8+c_8)-a_3-c_3})\ ,$$

The geometric hierarchy of the mass ratios in each quark sector
suggests the further equalities
$$a_8+b_8=a_3+b_3\ ;\qquad a_8+c_8=a_3+c_3\ .\eqno(aaa)$$
Agreement with experimental information on the quark mass ratios
dictates the following
$$2(a_8+c_8)=y(a_8+b_8)\ .\eqno(geom)$$
In addition the mixing angle relation
$$V_{us}=\sqrt{m_d\over m_s}\ ,\eqno(rel1)$$
is satisfied provided that
$$c_3=a_3\ ,\eqno(rel2)$$
if $y>1$, and
$$c_3=(2y-1)a_3\ ,\eqno(rel3)$$
if $0<y<1$.We also find that
$$V_{cb}={\cal O}(V_{us}^{3a_8-a_3\over 2a_3})\ ,\eqno(rel4)$$
from which we may deduce that
$$3a_8=5a_3\ .\eqno(ttt)$$
Comparison with the data gives us six equations among seven
unknown. The last unknown  is $y$.  Until we know the origins of the
scales and
of the expansion parameters, we cannot fix the values of $\lambda_d$ and
of $\lambda_u$ in terms of observables. We note, for example, the
interesting case $y=2$, corresponding to $\lambda^{}_u=\lambda_d^2$,
yields $b_i=c_i$, which suggests an $SU(2)_R$ symmetry.
It is quite remarkable that this simple idea is in agreement with the
present data, and even predicts one successful relation among the CKM
matrix elements \(bbb).

\subhead{3.~\bf Lepton Masses and Mixing Angles}
\taghead{3.}
An analysis  akin to that in the previous section yields the charged
lepton mass estimates
$${m_e\over m_\tau}={\cal
O}(\lambda_e^{3(d_8+e_8)+d_3+e_3})\ ; \qquad {m_\mu\over m_\tau}={\cal
O}(\lambda_e^{3(d_8+e_8)-d_3-e_3})\ .\eqno(kkk)$$
Geometric hierarchy of the charged lepton mass ratios implies that
$$d_8+e_8=d_3+e_3\ ,\eqno(kkl)$$
There are no mixing angles if the neutrinos are massless.
Below, we generalize the Froggatt-Nielsen analysis to massive
neutrinos, without assuming any extra symmetry[\cite{DLLRS}].
We do this by adding right-handed neutrinos to the MSSM
in order to generate masses for the neutrinos via the
$``$see-saw" mechanism[\cite{seesaw}].

Let us assume that the low energy chiral remnants of the primal soup
come from ${\bf 27}$ representations of $E_6$. This representation
carries two fields with no electroweak quantum numbers. One is an
$SO(10)$ singlet, as we can see from the decomposition
${\bf 27}={\bf 16}\oplus {\bf 10}\oplus {\bf 1}.$
The other is an $SU(5)$ singlet which lives in the spinor
representation of $SO(10)$
${\bf 16}={\bf\overline 5}\oplus{\bf 10}\oplus{\bf 1}.$
This same field is part of an isodoublet under the right-handed
$SU(2)_R$ inside $SO(10)$
${\bf 16}=({\bf 2},{\bf 1},{\bf\overline 3}^c\oplus {\bf 1}^c)\oplus
({\bf 1},{\bf 2},{\bf 3}^c\oplus {\bf 1}^c).$
These two neutrino fields are not so $``$ino" as they are assumed to
be very massive. With two fields, the Majorana mass matrix is
$$\pmatrix{0&m_1~m_2\cr
m_1\atop m_2&{\cal M}_0\cr}\ ,$$
where ${\cal M}_0$ is a $2\times 2$ symmetric matrix, and $m_{1,2}$
are the usual $\Delta I_w=1/2$ mass entries of electroweak order. Let
the eigenvalues of ${\cal M}_0$ be $M_1$ and $M_2$. We can
go to a basis where $M_0$ is diagonal,  in which $m_{1,2}$ are rotated
into $\hat m_{1,2}$, yielding the light eigenvalue
$$m_\nu={\hat m^2_1M_2-\hat m^2_2M_1\over M_1M_2}\ .$$
Thus if $M_1< M_2$, it becomes just $\hat m_1^2/M_1$, so that it is
the lighter of the singlet neutrinos that enters in the light neutrino mass.
This assumes that the $\hat m_i$ are of the same order of magnitude,
themselves much smaller than $M_1$.

Thus in the following we assume only one right-handed neutrino per
family, and leave the more complicated analysis to others. Assume that
we have three such fields, $\overline N_i$, each carrying X-charge.
The superpotential now contains the new interaction  terms

$$L_i{\overline N}_jH_u\left({\theta\over m_\nu}\right)^{p_{ij}}
+m_0{\overline N}_i{\overline N}_j\left({\theta\over m_0}\right)^{q_{ij}}
\ ,\eqno(kll)$$
multiplied by couplings of order one, and where $m_0$
is some mass of the order of the GUT scale or string scale.
In analogy with the quark and
charged lepton sectors, we assume that $p^{}_{33}=0$, so that there
is only the tree-level coupling for the third family.
Call the X-charges of the right-handed neutrinos $f_0,f_3,f_8$, so
that at tree-level

$$d_0+f_0=2(d_8+f_8)\ .\eqno(lll)$$
All  Yukawa couplings satisfy conservation of X,
relating $q_{ij}$ and $p_{ij}$ to the X-charges of the fields.
For three families, the $6\times 6$ Majorana mass matrix is of the
form
$$\pmatrix {0&{\cal M}\cr
{\cal M}^T&{\cal M}_0\cr}\ .$$
In the above ${\cal M}$ is the $\Delta I_w=1/2$ mass matrix with entries
not larger
than the electroweak breaking scale, and ${\cal M}_0$ is
the unrestricted $\Delta I_w=0$
mass matrix. Assuming that the order of magnitude of the $\Delta
I_w=0$ masses is much larger than the electroweak scale, we
obtain the generalized $``$see-saw'' mechanism.

The calculation of the light neutrinos masses and mixing angles
proceeds in two steps.
Let  $U^{}_0$ be the matrix which diagonalizes ${\cal M}_0$, that is
$${\cal M}_0=U^{}_0D^{}_0U_0^T\ ,\eqno(hhh)$$
where $D_0$ is diagonal. Then
in terms of $D_\nu$, the $3\times 3$ eigenvalue
matrix for the light neutrinos, and $U_\nu$ be their  mixing matrix,
we have
$${\bf Y}^\prime_e\equiv U^{}_\nu D^{}_\nu U_\nu^T
=-{\cal M}^\prime{1\over D_0}{\cal M}^{\prime T}\ .\eqno(ggg)$$
In the $``$see-saw" limit, the matrices $U_0$ and $U_\nu$ are unitary, so
that
$${\cal M}^\prime={\cal M}U_0^*\ .\eqno(fff)$$
The orders of magnitude of the heavy neutrino mass matrix  are
$${\cal M}_0=m_0{\cal O}\pmatrix{\lambda_0^{2(f_0+f_3+f_8)}&
\lambda_0^{2(f_0+f_8)}&
\lambda_0^{2f_0+f_3-f_8}\cr
\lambda_0^{2(f_0+f_8)}&\lambda_0^{2(f_0-f_3+f_8)}&\lambda_0^{2f_0-f_3-f_8}\cr
\lambda_0^{2f_0+f_3-f_8}&\lambda_0^{2f_0-f_3-f_8}
&\lambda_0^{2(f_0-2f_8)}}\ .$$
Its diagonalization yields the three eigenvalues
$$M_1=m_0{\cal O}(\lambda_0^{2(f_0+f_3+f_8)})~<~ M_2=m_0{\cal
O}(\lambda_0^{2(f_0-f_3+f_8)})~<~ M_3=
m_0{\cal O}(\lambda_0^{2(f_0-2f_8)})\ .$$
We have assumed for simplicity that the charges satisfy the
inequalities
$$f_0>2f_8\ ,\qquad 3f_8>f_3>0\ ,\eqno(ineq)$$
corresponding to $M_1<M_2<M_3$. The diagonalizing matrix is
$$U_0=m_0{\cal O}\pmatrix{1&
\lambda_0^{2f_3}&
\lambda_0^{3f_8+f_3}\cr
\lambda_0^{2f_3}&1&\lambda_0^{3f_8-f_3}\cr
\lambda_0^{3f_8+f_3}&\lambda_0^{3f_8-f_3}
&1}\ .$$
The electroweak breaking mass yields the matrix
$$
{\cal M}=m{\cal O}\pmatrix{
\lambda_\nu^{3(d_8+f_8)+d_3+f_3}
&\lambda_\nu^{3(d_8+f_8)+d_3-f_3}
&\lambda_\nu^{3d_8+d_3}\cr
\lambda_\nu^{3(d_8+f_8)-d_3+f_3}
&\lambda_\nu^{3(d_8+f_8)-d_3-f_3}
&\lambda_\nu^{3d_8-d_3}\cr
\lambda_\nu^{3f_8+f_3}
&\lambda_\nu^{3f_8-f_3}
&1\cr}
\ ,\eqno(sss)$$
where $\lambda_\nu$ is the expansion parameter, and $m$ is a mass of
electroweak breaking size.
If we let $\lambda_0=\lambda_\nu^z$, with $z>0$. When $z\ge 1$, we find that
$${\bf Y}_e^\prime={m^2\over M_1}{\cal O}(\lambda_\nu^{6f_8+2f_3})
{\cal O}\pmatrix{\lambda_\nu^{6d_8+2d_3}
&\lambda_\nu^{6d_8}
&\lambda_\nu^{3d_8+d_3}\cr
\lambda_\nu^{6d_8}&
\lambda_\nu^{6d_8-2d_3}
&\lambda_\nu^{3d_8-d_3}\cr
\lambda_\nu^{3d_8+d_3}
&\lambda_\nu^{3d_8-d_3}
&1\cr}
\ ,\eqno(ttu)$$
is the matrix whose eigenvalues yield the light neutrino masses, and
their mixing angles. It is diagonalized by the unitary matrix
$$U_\nu={\cal O}\pmatrix{1&
\lambda_\nu^{2d_3}&
\lambda_\nu^{3d_8+d_3}\cr
\lambda_\nu^{2d_3}&1&\lambda_\nu^{3d_8-d_3}\cr
\lambda_\nu^{3d_8+d_3}&\lambda_\nu^{3d_8-d_3}
&1}\ .\eqno(diaglight)$$
The light neutrino masses are then
$$\eqalign{m_{\nu_1}&={m^2\over M_1}{\cal
O}(\lambda_\nu^{2(3f_8+3d_8+f_3+d_3)})\
,\cr
m_{\nu_2}&={m^2\over M_1}{\cal O}(\lambda_\nu^{2(3f_8+3d_8+f_3-d_3)})\ ,\cr
m_{\nu_3}&={m^2\over M_1}{\cal O}(\lambda_\nu^{2(3f_8+f_3)})\
.\cr}\eqno(light)$$
In order to obtain the  mixing matrix which appears in the charged lepton
current, we must fold this matrix with that which diagonalizes the
charged lepton masses. If we let $\lambda^{}_\nu=\lambda_e^w$, with
$w>1$, the result is

$${\cal U}_\nu={\cal O}\pmatrix{1&
\lambda_e^{2d_3}&
\lambda_e^{3d_8+d_3}\cr
\lambda_e^{2d_3}&1&\lambda_e^{3d_8-d_3}\cr
\lambda_e^{3d_8+d_3}&\lambda_e^{3d_8-d_3}
&1}\ .$$
When $0<w<1$, the matrix has the same form with $\lambda_e$ replaced
by $\lambda_\nu$.
It is similar to the CKM matrix.
The mixing in the charged lepton current was first proposed by Maki,
Nakagawa and Sakata in 1962, so we call it the MNS
matrix[\cite{NAGOYA}].
We note that its elements satisfy
$$V_{e\nu_\mu} V_{\mu\nu_\tau}\sim V_{e\nu_\tau}\ .\eqno(leptmix)$$
It may be that $\lambda_e=\lambda_d$ and $\lambda_u=\lambda_\nu$,
since they have the same quantum numbers, implying $w=y$.
We also have the relations
$${m_{\nu_1}\over m_{\nu_2}}\approx (V_{e\nu_\mu})^w\ ;
\qquad{m_{\nu_2}\over m_{\nu_3}}\approx (V_{\mu\nu_\tau})^w\
,\eqno(neutes)$$
valid only when $w>1$. When $0<w<1$ the exponents in these relations
is one.
In this  analysis, the lepton mixing matrix has the same
structure as the CKM matrix. We have assumed a simple set of
inequalities among the charges, to provide an example of our method.
When $0<z<1$, the forms of the neutrino masses are the same except
that $f_{3,8}$ appear multiplied by $z$. The mixing matrix is
unchanged.

Unlike quark masses and mixing, we have little solid experimental
information on the values of these parameters. The most compelling
evidence for neutrino masses and mixings come from the MSW
interpretation of the deficit observed in various solar neutrino
fluxes. In this picture, the electron neutrino mixes with another
neutrino (assumed here to be the muon neutrino)
with a mixing angle $\theta_{12}$ such that
$$m^2_{\nu_1}-m^2_{\nu_2}\approx 7\times 10^{-6}~{\rm eV^2}\ ;\qquad
\sin^22\theta_{12}\approx 5\times 10^{-3}\ .\eqno(solar)$$

The other piece of evidence comes from the deficit of muon neutrinos
in the collision of cosmic rays with the atmosphere.
If taken at face value, these suggest that the muon neutrinos
oscillate into another species of neutrinos, say $\tau$ neutrinos,
with a mixing angle $\theta_{23}$, and masses such that
$$m^2_{\nu_2}-m^2_{\nu_3}\approx 2\times 10^{-2}~{\rm eV^2}\ ;\qquad
\sin^22\theta_{23}\ge .5  \ .\eqno(atmos)$$
Fitting the parameters coming from the solar neutrino data is rather
easy, suggesting that
$$V_{e\nu_\mu}\sim\lambda_e^{2d_3}\sim \lambda^2\ ,$$
together with $m_{\nu_2}\approx 1$ meV. However it is not so easy to
understand the atmospheric neutrino data. These imply
$$V_{\mu\nu_\tau}\sim \lambda_e^{3d_8-d_3}={\cal O}(1)\ .$$
The relations \(neutes) then suggest that
$w$ has to be large. For example the
value $\theta_{23}\sim
{\pi\over 9}$ for which $\sin^22\theta_{23}=.34$,
yields $m_{\nu_2}/ m_{\nu_3}\sim .01$ , for $w=4$. This gives
$m_{\nu_3}\approx .1$ eV, which marginally reproduces the $``$data", and
fixes the lightest neutrino mass to $m_{\nu_1}\approx 10^{-13}$ eV!
The heaviest neutrino weighs one tenth of an eV, not enough to be of use
for structure formation. Perhaps there are more light neutrals, coming
from the extra neutral leptons in
each $E_6$ or from  {\it  end T} in string compactification.

Generically, though, it is difficult to understand
mixing angles of order one, as suggested by the atmospheric neutrino
data. The existence of only small mixing angles in
the quark sectors suggests either
that the interpretation of the atmospheric neutrino data is premature,
or that there is fine tuning in the neutrino matrices[\cite{RK}].
\vskip .5cm
\subhead{4.~\bf Anomalies}
\taghead{4.}
The X family symmetry is in general anomalous.  The three chiral families
contribute to the mixed gauge anomalies as follows
$$\eqalignno{C_3&=3(2a_0^{}+b_0^{}+c_0^{})\ ,&(anom3)\cr
C_2&=3(3a_0^{}+d_0^{})\ ,&(anom2)\cr
C_1&=a_0^{}+8b_0^{}+2c_0^{}+3d_0^{}+6e_0^{}\ .&(anom1)\cr}$$
The subscript denotes the gauge group of the Standard Model, {\it i.e.}
$1\sim U(1)$, $2\sim SU(2)$, and $3\sim SU(3)$.
The X-charge also has a mixed gravitational anomaly, which is simply
the trace of the X-charge,
$$C_g=3(6a_0^{}+3b_0^{}+3c_0^{}+2d_0^{}+e_0^{}+f_0^{})-x+C_g^\prime\
,\eqno(anomg)$$
where $C_g^\prime$ is the contribution from the particles that do not
appear in the model we are discussing.
One must also account for the mixed $YXX$ anomaly, given by
$$C_{YXX}=6(a_0^2-2b_0^2+c_0^2-d_0^2+e_0^2)+4A_T\ ,\eqno(anomixed)$$
with the texture-dependent part given by
$$A_T=(3a_8^2+a_3^2)-2(3b_8^2+b_3^2)+
(3c_8^2+c_3^2)-(3d_8^2+d_3^2)+(3e_8^2+e_3^2) \ .\eqno(AT)$$
The last anomaly coefficient is that of the X-charge itself, $C_X$,
the sum of the cubes of the X-charge.
Extra particles with chiral X-charge other than those in the minimal
model, will contribute to both $C^\prime_g$ and $C_X$.

These anomaly coefficients can be related to combinations of quarks
and lepton masses. The reason is that  the
X-charge of the determinant in each charge sector is {\it
independent} of the texture coefficients that distinguish
between the two lightest families. We set
$$\det {\bf Y}_u\sim
y_t^3 {\cal O}(\lambda_u^{U})\ ,\qquad \det {\bf Y}_d\sim y_b^3
{\cal O}(\lambda_d^{D})\ ,\qquad \det {\bf Y}_l\sim y_\tau^3
{\cal O}(\lambda_e^{E})\ ,\eqno(detall)$$
where
$$U\equiv
6(a_8+b_8)\ ,\ D\equiv 6(a_8+c_8)\ ,\  E\equiv 6(d_8+e_8)\ .$$

Since the down and lepton matrices have the same quantum numbers, and
couple to the same Higgs, we may assume they have the same expansion
parameter, $\lambda_d=\lambda_e$. In that case we can relate the
products of the down quark masses to that of the leptons (assuming
$y_b=y_\tau$)
$${ m_dm_sm_b\over m_em_\mu m_\tau}\sim
{\cal O}(\lambda_d^{(D-E)})\ . \eqno(detratio)$$

{}From the tree-level Yukawa couplings to the third family expressed
through \(Yuky), we can write
combinations of anomaly coefficients in terms of the family-dependent
charges
$$\eqalign{C_1+C_2-{8\over 3}C_3&=12(d_8+e_8-a_8-c_8)=2(E-D)\ ,\cr
C_3&=6(2a_8+b_8+c_8)=U+D\ .\cr}\eqno(anotext)$$
These  allow us to relate the anomaly
coefficients to the ratio of products of quark and lepton masses
\(detratio), (assuming $y_b=y_\tau$),
$$  {m_dm_sm_b \over m_em_\mu m_\tau}\sim {\cal
O}(\lambda_d^{-(C_1+C_2-8/3C_3)/2})\ .\eqno(detrat)$$
Compatibility with the extrapolated data requires the exponent to vanish
$$C_1+C_2-{8\over 3}C_3= 0\ ,\eqno(vanish)$$
which expressed in other variables, reads $E=D$.

\subhead{5.~\bf Green-Schwarz Cancellation of X Anomaly}
\taghead{5.}
If X is anomaly-free, then
$$C_1=C_2=C_3=0\ ,\qquad C^{}_g=0\ .\eqno(anofree) $$
The last equation is not constraining as there are likely more fields in
the theory with chiral X-charge. These are  consistent with
\(detrat), but the vanishing of $C_3$ contradicts our hypothesis that
all excess charges have the same sign. Indeed, using the tree-level
Yukawa relations \(Yuky), \(anom3), we see that
$$0=C_3=6(a_8+b_8)+6(a_8+c_8)\ ,$$
which is not consistent with our assumption that all excess charges are
positive. Hence we must rely on the Green-Schwarz mechanism.

String theories naturally contain all of the ingredients we need to
reproduce the Yukawa textures. They have an antisymmetric tensor
Kalb-Ramond field which in four dimensions is the Nambu-Goldstone boson
of an anomalous $U(1)$ that couples like an axion through a dimension
five term to the divergence of the anomalous current. Its anomalies are
cancelled by the Green-Schwarz mechanism[\cite{GS}]. Under a chiral
transformation, this term is capable of soaking up certain anomalies, by
shifting the axion field, provided that they appear in commensurate
ratios
$${C_i\over k_i}={C_X\over k_X}={C_g\over k_g}\ ,\eqno(proportion)$$
where the $k_i$ are the Kac-Moody levels. They need to be integers only
for the non-Abelian factors.

In superstring theories, this $U(1)$ is broken spontaneously slightly
below the string scale. The scale is set by the charge content of the
theory[\cite{break}]. It follows that singlets with masses protected by
X can still be very massive, and not appear in the effective low-energy
theory.

This chiral U(1) X-charge can fix the value of the Weinberg
angle, without the use of a grand unified group, as remarked by Ib\` a\~
nez[\cite{Ib}. More recently, Ib\` a\~
nez and Ross[\cite{IR}] applied it to the determination of symmetric
textures when the field $\theta$ is vector-like.

In superstring theories, the non-Abelian gauge groups have the same
Kac-Moody levels. For Green-Schwarz cancellation, it means that
$$C_2=C_3\qquad {\rm or}\qquad  d_0^{}=b_0^{}+c_0^{}-a_0^{}\ .\eqno(kac)$$
After this very generic requirement, we see that equation \(detrat)
reduces to
$$  {m_dm_sm_b \over m_em_\mu m_\tau}\sim {\cal
O}(\lambda_d^{-(C_1-5/3C_2)/2})\ ,\eqno(detrat2)$$
valid whenever $\theta$ is chiral. Since the right-hand side is of order
one, it means that  the exponent vanishes, so that in models
with an {\it ab initio} $\mu$ term, we {\it deduce} that
$$C_1={5\over 3}C_2\ .\eqno(good)$$
However the gauge coupling constants at string unification scale with
the anomaly coefficients, so that
$${C_1\over C_2}={g_1^2\over g_2^2}\ ,\eqno(weinberg)$$
which fixes the Weinberg angle to the value
$$\sin^2\theta_w={3\over 8}\ ,$$
at the string scale, the canonical GUT value, but without the excess
baggage of these theories! This is a strong hint that the $N=1$ model does
indeed come from superstrings!

We note that the mixed gravitational anomaly is exactly along the
anomaly-free combination of baryon minus lepton numbers, $B-L$. In
fact the most general X-charge can contain an arbitrary mixture along
$B-L$, but this is already taken into account by our general
parametrization.
In superstring models, the Green-Schwarz mechanism extends to the mixed
gravitational anomaly so that
$${C_g\over C_3}={k_g\over k_3}=\eta\ .$$
where $\eta$ is a normalization parameter; in the simplest
level-one models, it is equal to $12$. In general, however,
$$C_g=\eta(U+E)\ .\eqno(gsmixed)$$
The family independent X-charges are seen to depend only on two
parameters, $E$, and $U$, assuming we know the normalization $\eta$.

Starting from very simple generic assumptions we are able to reproduce
the data and even determine the Weinberg angle in terms of the ratio
of quark and lepton masses. Our analysis, applied to neutrino
masses, shows that it is ackward to accomodate both solar neutrino
and atmospheric neutrino data.

I would like to thank Professors Vinet for inviting me to contribute
to this anniversary volume. My interactions with the Centre de
Recherches Math\' ematiques started in 1973, when Professor J. Patera
showed me his treasure trove on exceptional groups, at a time where
they were but curiosities. They are now part of the physics
vocabulary, as in the heterotic string and grand unification.
I wish to thank M. Booth for reading this manuscript.
This work was supported in part by the United States
Department of Energy under grant DE-FG05-86-ER40272

\references

\refis{unification} P.~Langacker, in Proceedings of the
PASCOS90 Symposium, Eds.~P.~Nath
and S.~Reucroft, (World Scientific, Singapore 1990).

\refis{FN} C.~Froggatt and H.~B.~Nielsen \np B147, 277, 1979.

\refis{reviews}
For reviews, see H.~P.~Nilles,  \prpts 110, 1, 1984 and
H.~E.~Haber and G.~L.~Kane, \prpts 117, 75, 1985.

\refis{gut}
J.~C.~Pati and A.~Salam,
\pr D10, 275, 1974;
H.~Georgi and S.~Glashow,
\prl 32, 438, 1974;
H.~Georgi, in {\it Particles and Fields-1974}, edited by C.E.Carlson,
AIP Conference Proceedings No.~23 (American Institute of Physics,
New York, 1975) p.575;
H.~Fritzsch and P.~Minkowski,
\journal Ann.~Phys.~NY, 93, 193, 1975;
F.~G\" ursey, P.~Ramond, and P.~Sikivie,
\pl 60B, 177, 1975.

\refis{btau} H.~Arason, D.~J.~Casta\~no, B.~Keszthelyi, S.~Mikaelian,
E.~J.~Piard, P.~Ramond, and B.~D.~Wright,
\prl 67, 2933, 1991;
A.~Giveon, L.~J.~Hall, and U.~Sarid,
\pl 271B, 138, 1991.

\refis{RRR}P. Ramond, R.G. Roberts and G. G. Ross, \np B406, 19, 1993.


\refis{Ib}L. Ib\'a\~nez, \pl B303, 55, 1993.

\refis{IR}L. Ib\'a\~nez and G. G. Ross, \pl B332, 100, 1994.

\refis{unification}
U.~Amaldi, W.~de Boer, and H.~Furstenau,
\pl B260, 447, 1991;
J.~Ellis, S.~Kelley and D.~Nanopoulos,
\pl 260B, 131, 1991;
P.~Langacker and M.~Luo,
\pr D44, 817, 1991.

\refis{hall} S. Dimopoulos, L. Hall, S. Raby, and G. Starkman, \pr
D49, 3660, 1994.

\refis{LNS} M. Leurer, Y. Nir, and N. Seiberg, \np B398, 319, 1993.

\refis{GS} M. Dine, N. Seiberg, and E. Witten, \np B289, 585, 1987;
J. Atick, L. Dixon, and A. Sen, \np B292, 109, 1087;
M.  Dine, I. Ichinoise, and N. Seiberg, \np B293, 253, 1987.

\refis{break} A. Font, L.E. Ib\' a\~ nez, H. P. Nilles, and F. Quevedo,
\np B307, 109, 1988; \pl B210, 101, 1988;
J. A. Casas, E. K. Katehou, and C. Mu\~ noz, \np B317, 171, 1989;
J. A. Casas, and C. Mu\~ noz, \pl B209, 214, 1988; \pl B214, 63, 1988;
A. Font, L.E. Ib\' a\~ nez, F. Quevedo, and A. Sierra, \np B331, 421,
1990.

\refis{NAGOYA}Z. Maki, M. Nakagawa, and S. Sakata,\ptp 28, 247, 1962.




\refis{seesaw}M. Gell-Mann, P. Ramond, and R. Slansky in Sanibel
Talk,
CALT-68-709, Feb 1979, and in {\it Supergravity} (North Holland,
Amsterdam 1979). T. Yanagida, in {\it Proceedings of the Workshop
on Unified Theory and Baryon Number of the Universe}, KEK, Japan,
1979.

\refis{wolf} L. Wolfenstein, \prl  51, 1945, 1983.

\refis{BLR} P. Bin\'etruy, S. Lavignac, and P. Ramond, in preparation.

\refis{BR} P. Bin\'etruy and P. Ramond, LPTHE-ORSAY 94/115,
UFIFT-HEP-94-19, to appear in {\it Phys Lett B}.

\refis{PAPAG} E. Papageorgiou, $``$Yukawa Textures from an extra $U(1)$
Symmetry", Orsay Preprint, LPTHE Orsay 40/94.%

\refis{DLLRS} H. Dreiner, G. Leontaris, S. Lola, G. Ross, and C.
Scheich, \np B436, 461, 1995.

\refis{JS} V. Jain, and R. Shrock, Stony Brook Preprint, Dec 1994.

\refis{DPS} E. Dudas, S. Pokorski, and C. Savoy, Saclay preprint, MArch
1995

\refis{NIR} Y. Nir, private communication.

\refis{RK} S.P. Rosen and W. Kwong, Univ of Texas at Arlington
preprint, UTAPHY-HEP-13 (1995).

\endreferences\endit
\end